\documentclass[ijds, nonblindrev]{informs-ijds}

\OneAndAHalfSpacedXI

\usepackage{natbib}
 \bibpunct[, ]{(}{)}{,}{a}{}{,}%
 \def\bibfont{\small}%

\usepackage{hyperref}
\usepackage{booktabs}

\TheoremsNumberedThrough     
\ECRepeatTheorems

\EquationsNumberedThrough    

\begin{document}



\RUNTITLE{Neuroadaptive electroencephalography}
\RUNAUTHOR{Pedro F. da Costa et al.}

\TITLE{Neuroadaptive electroencephalography:\break a proof-of-principle study in infants
}

\ARTICLEAUTHORS{%
\AUTHOR{Pedro F. da Costa}
\AFF{Centre for Brain and Cognitive Development, Department of Psychological Science, Birkbeck, University of London, Henry Wellcome Building, Malet Street, London WC1E 7HX, United Kingdom} 
\AFF{Department of Neuroimaging, Institute of Psychiatry, Psychology and Neuroscience, King's College London, de Crespigny Road, London SE5 8AB, United Kingdom \EMAIL{pedro.ferreira\_da\_costa@kcl.ac.uk}}
\AUTHOR{Rianne Haartsen}
\AFF{Centre for Brain and Cognitive Development, Department of Psychological Science, Birkbeck, University of London, ToddlerLab, Malet Street, London WC1E 7HX, United Kingdom \EMAIL{rhaart01@mail.bbk.ac.uk}} 
\AUTHOR{Elena Throm}
\AFF{Centre for Brain and Cognitive Development, Department of Psychological Science, Birkbeck, University of London, ToddlerLab, Malet Street, London WC1E 7HX, United Kingdom \EMAIL{ethrom01@mail.bbk.ac.uk}} 
\AUTHOR{Luke Mason}
\AFF{Centre for Brain and Cognitive Development, Department of Psychological Science, Birkbeck, University of London, Henry Wellcome Building, Malet Street, London WC1E 7HX, United Kingdom \EMAIL{l.mason@bbk.ac.uk}}
\AUTHOR{Anna Gui}
\AFF{Centre for Brain and Cognitive Development, Department of Psychological Science, Birkbeck, University of London, Henry Wellcome Building, Malet Street, London WC1E 7HX, United Kingdom \EMAIL{agui01@mail.bbk.ac.uk}}
\AUTHOR{Robert Leech}
\AFF{Department of Neuroimaging, Institute of Psychiatry, Psychology and Neuroscience, King's College London, de Crespigny Road, London SE5 8AB, United Kingdom \EMAIL{robert.leech@kcl.ac.uk}}
\AUTHOR{Emily J. H. Jones}
\AFF{Centre for Brain and Cognitive Development, Department of Psychological Science, Birkbeck, University of London, Henry Wellcome Building, Malet Street, London WC1E 7HX, United Kingdom \EMAIL{e.jones@bbk.ac.uk}}

} 

\ABSTRACT{%
A core goal of functional neuroimaging is to study how the environment is processed and represented in the brain. The mainstream paradigm involves concurrently measuring a broad spectrum of brain responses to a small set of environmental features preselected with reference to previous studies or a theoretical framework. As a complement, we invert this approach by allowing the investigator to record the modulation of a preselected brain response by a broad spectrum of environmental features. Our approach is optimal when theoretical frameworks or previous empirical data are impoverished or misleading: for example, the appropriate stimulus choice may be unclear when asking new questions or moving beyond the modal population in neuroimaging (heteronormative undergraduate students). By using a prespecified closed-loop design, the approach addresses fundamental challenges of reproducibility and generalisability in brain research. These conditions are particularly acute when studying the developing brain, where our theories based on adult brain function may fundamentally misrepresent the topography of infant cognition and where there are substantial practical challenges to data acquisition. 

Our methodology begins with a neural phenomenon of interest (e.g., an event-related neural response) and employs machine learning and real-time data analysis to map modulation of this neural feature across a space of experimental stimuli. Our method collects, processes and analyses EEG brain data in real-time; and uses a neuro-adaptive Bayesian optimisation algorithm to adjust the stimulus presented depending on the prior evidence obtained for a given participant. Unsampled stimuli can be interpolated by fitting a Gaussian process regression along the dataset. As a proof-of-principle with a known endpoint, we show that our method can automatically identify the face of the infant’s mother through online recording of their Nc brain response to a face continuum. We can retrieve model statistics of individualised responses for each participant, opening the door for early identification of atypical development. This approach has substantial potential in infancy research and beyond for improving power and generalisability of mapping the individual cognitive topography of brain function.

}%


\KEYWORDS{Neuroadaptive approach, EEG, infants, Bayesian optimization, face processing, generative models}

\maketitle

%


\section{Introduction}

Understanding how the brain processes and represents the physical and social environment is one of the fundamental goals of functional neuroimaging.  Decades of research have yielded a range of methodologies for studying the electrical activity (electroencephalography/EEG), magnetic activity (magnetoencephalography; MEG) and oxygenated haemoglobin changes (functional magnetic resonance imaging/MRI and near-infrared spectroscopy/NIRS) that are associated with neuronal activity in the human brain. Further, cognitive neuroscience has generated a rich tapestry of neural metrics suitable for assessing response to environmental stimulation.  These include measurement of task-induced changes in oxygenated haemoglobin concentrations in spatially-defined regions or networks (fMRI/NIRS, and timing-defined event-related neural potentials or oscillations (EEG/MEG). These metrics were identified through experiments in which a researcher measured a broad spectrum of brain responses (e.g., whole-brain EEG or fMRI) to a small pre-selected stimulus set. Often, the selected stimuli are important environmental cues like faces \citep{Tsao2006, Tsao2008a}, or basic auditory or visual features designed to represent the building blocks of perception (like checkerboards or tones). Such research has yielded many critical insights into metrics relevant to brain function \citep{Kropotov2009}.

Despite this progress, there are increasing questions over the value of traditional stimulus-driven methods for understanding the ‘meaning’ of these brain metrics (i.e., what dimensions they represent); and for studying how brain function differs in populations who are not the modally-studied groups of heteronormative White western young adults. First, understanding what each brain metric ‘means’ involves defining the range of stimuli by which each brain metric is modulated; doing this sequentially through separate experiments that each focus on one or two stimuli is slow and inefficient. Second, stimulus selection is often guided by theory or previous empirical study; but the replication crisis \citep{Ioannidis2014, Collaboration2015} and the overwhelming focus on a narrow subset of the world’s population in the neuroimaging literature \citep{Westfall2017} means that trying to select the right stimuli to study individual differences or brain function in broader populations may be little better than guessing. The substantial analytic flexibility afforded by allowing post-experiment analysis of brain data creates an overwhelming risk of false positives and can only be partially addressed by written pre-registration \citep{Nosek2015}. Finally, the traditional approach embeds a deficit model in our approach to studying individuals with neurodevelopmental or psychiatric disorders. A typical research project in this area involves studying how people with neurodevelopmental or psychiatric disorders respond differently to a stimulus selected based on normative preferences. Reframing our work within a neurodiversity framework \citep{Singer2009} prompts us to ask not how people’s brains respond differently to stimuli that ‘typical’ brains prefer, but to identify what stimuli are preferred by people whose brains work differently.  Longer-term, this may prove more fruitful for the design of individualised interventions that build on strengths, rather than aim to address weaknesses.

To address these problems, Leech and colleagues developed a complementary approach that inverts the traditional experimental paradigm \citep{Lorenz2018c}. Instead of preselecting one or two stimuli and measuring a broad spectrum of brain responses, neuroadaptive Bayesian optimisation is a method through which the experimenter selects one or two brain responses and measures how they are modulated by a broad spectrum of environmental stimuli in a real-time closed-loop design. This method has been used successfully to study fMRI responses in the frontal cortex in neurotypical adults \citep{Lorenz2018c} and to identify cognitive difficulties in stroke patients \citep{Lorenz2021}. Here, we extend this approach to EEG and the study of infant brain function. EEG is a particularly fruitful method for real-time analysis because its high temporal resolution allows rapid feedback loops of response-guided stimulus selection. Further, the problems of traditional approaches are particularly acute when studying the developing brain, where the cognitive topography is likely to be substantially different than in adulthood and where practical challenges of working with infants make data collection slow and difficult. However, without studying the brain as it develops, we cannot move beyond studying the correlation between environmental features and brain response. To understand the mechanisms of causation through which the environment is represented and shaped we must study change over developmental time in how the brain processes information and controls behaviour. Insights from translational work in animal models and computational modelling approaches to studying learning are also most likely to be effectively mapped onto preverbal infants, where common mechanisms are most likely to be conserved. Thus, in the present study we present a proof-of-principle of the use of neuroadaptive techniques to study the developing brain.

As a test case, we selected a stimulus space in which we could make a strong prediction about individual-level brain ‘preferences’. Specifically, we utilised a face space within which the face of the infant’s mother was positioned. If the algorithm can automatically converge to identify the infant’s mother based on real-time closed-loop analysis of their brain activity, this provides a strong proof-of-principle that the technique can be used to study how a particular brain metric reacts to a broad environment. Notably, this is different to a typical BCI approach in which an algorithm would first be trained to distinguish (for example) two faces on the basis of multimodal brain data; here, the neural feature is selected based on the previous literature for greater interpretability.  In this case, we selected the negative central (Nc) event-related potential response because of its demonstrated links to attention \citep{Richards2010} and its modulation by face familiarity (including differentiation between mother and stranger \citep{DeHaan1999, Luyster2014}. The infant Nc (negative component) is a negative deflection occurring between 300 and 800ms at the frontal midline after the stimulus onset. Previous studies showed that the amplitude of the Nc was larger (more negative) in response to the mother’s face \citep{DeHaan1999}, likely reflecting elevated attention. These studies use the traditional approach of preselecting two stimuli (mother and stranger) and analysing the resulting data with highly variable study-specific parameters (scalp location, window timing, peak or latency etc). To invert this paradigm, we generated facial stimuli based on the mothers and strangers’ faces and created a configuration space. We then presented one of the faces to the infants while measuring their EEG. Using real-time EEG analyses, we analysed their Nc amplitude response to the stimuli. The Nc responses were then forwarded to the neuroadaptive algorithm that predicted which face would elicit the optimal Nc response (i.e., larger Nc amplitude) in the individual infant. When the algorithm converged on the stimulus eliciting the optimal Nc response, the experiment was automatically terminated. In the following sections, we describe the method and proof-of-principle results.

\section{Materials and Methods}
\begin{figure}[t]
\begin{center}
\centerline{\includegraphics[width=\textwidth]{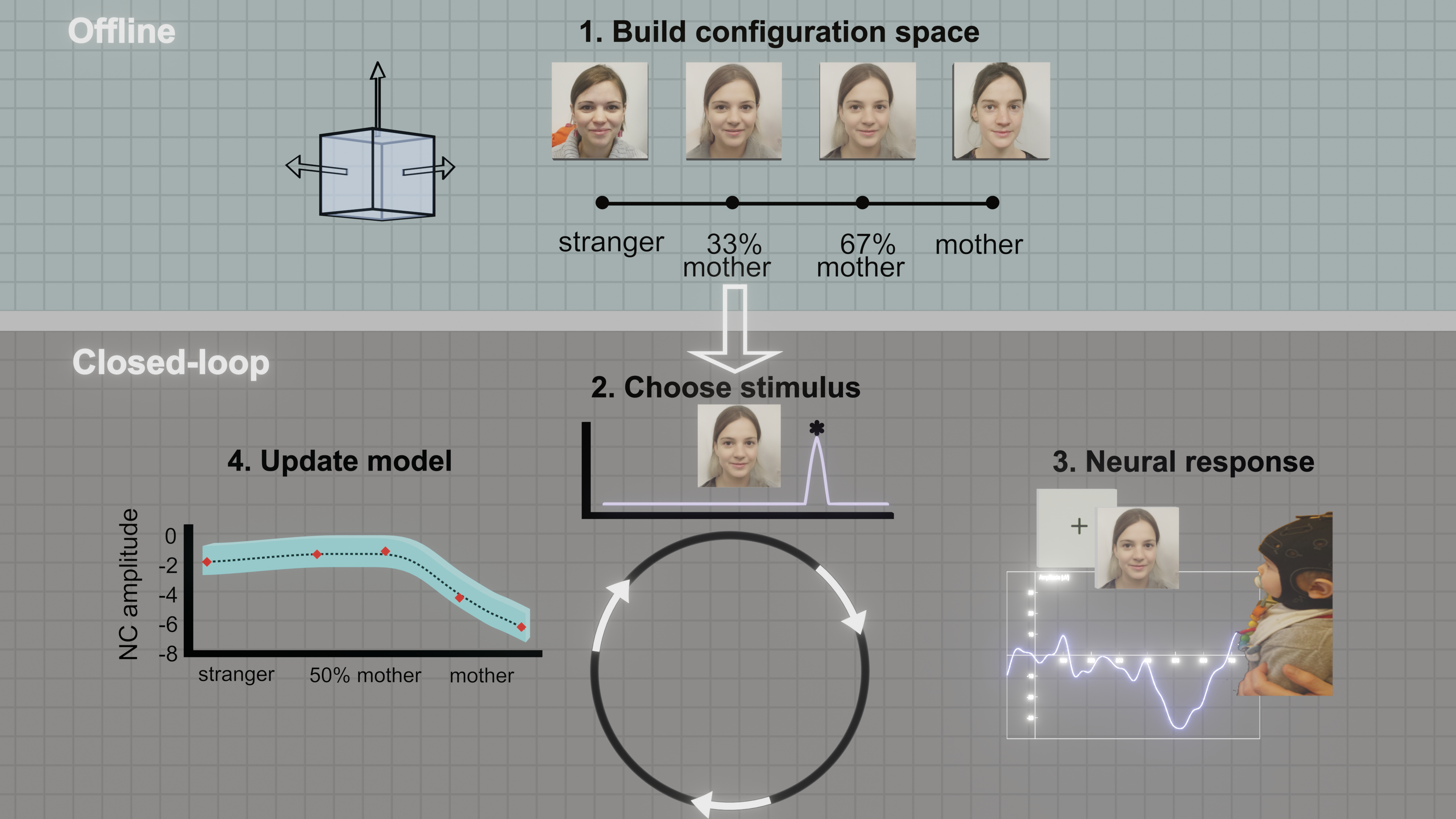}}
\caption{\textit{Schematic of the EEG neuroadaptive framework.} The first step is taken before the testing of a participant and consists of building the configuration space of stimuli (1.) - in our paradigm, mother-stranger interpolation. However, this could represent a stimulus space across a large number of dimensions. During testing, the framework is run in a closed-loop, where a stimulus in the configuration space is chosen to be sampled (2.). The stimulus is displayed to the participant and the EEG response is processed automatically to retrieve the target metric (3.) - in our paradigm, the Nc amplitude. A statistical model is built by fitting a Gaussian process to the sampled data (4.), guiding where to sample in the next iteration (2.).} \label{schematic}
\end{center}
\end{figure}

Our neuroadaptive method consists of multiple steps, outlined in the following sections. These are: generating stimuli and creating a configuration space (here a face space; \ref{gan}); recording and performing real-time analysis of the neural (EEG) responses to the stimuli (\ref{eeg}); using a sampling algorithm (Bayesian optimisation) across the space (\ref{bayesian}) and re-iterating steps 2 and 3 until a stopping criterion has been reached (\ref{stop}). This allows for a rapid prediction of the target EEG metric across the space despite sampling only a limited number of stimuli.  Figure \ref{schematic} presents an overview of these steps. In what follows we describe the development and choice of parameters for each step in more details. All scripts are available in an open-source repository \footnote{github.com/PedroFerreiradaCosta/NeuroadaptiveEEG/}.

\subsection{Generating stimuli and creating a configuration space}
\label{gan}
Our method maps the peak and topography of the modulation of the pre-selected neural response across a large stimulus space. As such, a continuous experimental space needs to be generated that can be characterised along one or more stimulus dimensions \citep{DaCosta2020a}. Since our test case involved faces (a common focus of cognitive neuroscience because of their importance to social function), to allow us to create a smooth space interpolating between faces of strangers and participant’s mothers, stimuli were artificially generated using StyleGAN2. StyleGAN2 is a state-of-the-art generative adversarial network (GAN), a deep learning algorithm for generative image modelling \citep{Karras2019a}. GANs are deep learning algorithms capable of generating artificial images barely indistinguishable from real images \citep{Goodfellow2014d}. They are trained in an adversarial setting, where two components try to optimise opposite functions in a minimax game. One component, the Generator, is trained to generate high-fidelity representations of the original data, while the second component, the Discriminator, is trained to identify real data from artificial data generated by the Generator. Progressively, the Generator learns the latent manifold of the real data distribution and can generate high-quality examples of artificial data, not present on the training set. 
One of the examples of successful trained GANs on a given data distribution is the generation of realistic faces \cite{Karras2019a}. The generative algorithm can create a large and diverse compendium of non-existing faces that are indistinguishable from real photos. Exploring the trained latent manifolds of generated faces allows us to create smooth spaces defined by relevant dimensions along which faces continuously vary. By using artificially generated faces we are not limited by a bounded dataset and we can maximise diversity of the generated faces, important to expanding research studies to diverse cultures and ethnicities. Furthermore, the generated images can be manipulated to create realistic representations of faces that progressively change across a given semantic dimension, which provides an alternative way to define a ‘space’ (e.g., changing a given face’s perceived age) \citep{Radford2016a}. Finally, as the generated faces do not depict real people, there is no risk of privacy infringement or limitations on proprietary datasets. An additional use of GANs is to project real images of faces to the latent manifold. This facilitates the manipulation of real images across selected semantic dimensions (i.e., a meaningful dimension in the latent manifold). This allows us to create a dimension involving a real person (here the infant’s mother) by following previous implementations of image encoding \citep{Bojanowski2018a}. Specifically, we project both the GAN output and the image we want to encode into a common feature space, encoded by a perceptual model – an intermediate layer of the image classifier VGG16 \citep{Simonyan2015a}. The latent codes, $d_{latent}$, are then optimised directly by gradient descent to try to minimise the perceptual loss. The perceptual loss is the difference between both projected images in the common feature space. Given the output of the feature space $F(I)$, where $I$ is the image input, the perceptual loss can be given by the Euclidean distance between the two projections: 
\begin{equation*}
L_{percept}(I_{1}, I_{2}) = {||F(I_{1}) - F(I_{2})||}^2 
\end{equation*}
The resulting latent code that minimizes the perceptual loss will be able to generate a similar recreation of the original photograph. As previous studies have demonstrated \citep{Radford2016a, DaCosta2020a}, the latent space can be exploited to control for different facial features of the artificially generated faces. We can progressively change a given aspect of the face while maintaining the other features mostly intact. To manipulate a given feature first, its characteristic distribution needs to be identified in the latent space of faces. This is possible by collecting generated images and labelling them according to a binary classification of the feature being analysed (e.g., male vs female). Then, a logistic regression is fitted to the latent codes of the two classes. The coefficient c of the fitted regression, a vector of the same dimensions of $d_{latent}$, will then navigate the linear direction that best controls the semantic separation between the two classes. The magnitude of change of the original generated image is controlled by a scalar, $m$, given by the following linear equation:
\begin{equation*}
d_{latent new} = d_{latent} + c * m
\end{equation*}
Finally, the same linear interpolation can be applied to any two images generated from the GAN. By linearly manipulating two latent codes from one to the other, we can obtain a continuous morphing of the faces, slowly changing their facial identity. These manipulations using the GAN latent space prove useful in neuroscientific research, as they allow for flexibility in facial features while maintaining realistic representations of the face. Instead of relying on categorical and discrete stimuli, these continuous variations can then be used for better assessment of how a given ERP varies. 

We do these manipulations of artificial faces using Google’s Colaboratory \footnote{github.com/PedroFerreiradaCosta/NeuroadaptiveEEG/blob/main/FaceShiftBirkbeck.ipynb}, an online Jupyter notebook environment that provides access to GPU computation, a requirement when doing inference with GANs. This module is run before the data collection and it is responsible for generating the stimuli that will compose the space being analysed. In this module, the experimenter will define the dimensionality of the space, the semantic direction of each dimension and the original stimulus under analysis. The semantic direction can be chosen from a list of available options: gender, age, emotion, yaw, roll, pitch, lip ratio, nose-ratio, eye-ratio, eye distance, eye-to-eyebrow distance, nose-to-mouth distance, mouth open/closed, eyes open/closed, nose tip position and interpolation between two different faces (the option used in the present study). In the present study we interpolate between the faces of the participant’s mother and a stranger. The original stimulus was an original photograph that is uploaded to the system; an alternative approach is to use a randomly generated artificial face. This configuration space builder is currently limited to facial stimuli, but it can be substituted by any stimuli that are predefined to model a configuration space to be analysed in real-time.

In our system, the configuration space is Euclidean, discrete and can be multi-dimensional. The number of dimensions, each of which should account for an independent variation of the stimuli, is only limited by the capability of the Bayesian optimisation to sample vast spaces, which is known to break down for spaces larger than 20 dimensions \citep{Frazier2018}. Each dimension should code a single and independent variation of the stimuli. This allows to disambiguate and eliminate confounders from the variation of the target metric along a given axis. The specific variations of stimuli being encoded per dimension should depend on the research questions being addressed. 

In summary, in this proof-of-concept infant paradigm, we focus on brain functioning during visual processing of familiar and unfamiliar faces: i.e., the mother’s face and a stranger’s face. The stranger’s face was the face of one of the researchers and was used across all infants. We used this stranger's face and a photograph of the mother of the infants as our extremes for our mother - stranger continuum. In the photographs, the mother and stranger had a neutral expression and the head was centred where the image was cropped at shoulder or clavicle height.  Using the StyleGAN2, we generated a continuous GAN latent space that represented the mother’s face linearly changing into the stranger’s face. This resulted in 10 additional, realistic images of faces, bringing the total of possible sampled stimuli to 12 (see Figure \ref{individual}).

\subsection{Recording and real-time analysis of the neural responses to the stimuli}
\label{eeg}

\subsubsection{Participants - }
Participants were recruited via a database of families interested in research. Infants were aged between 5 and 9 months and excluded if they had a family or personal history of epilepsy, were born preterm ($<$ 31 weeks gestational age), had a clinical diagnosis or a sensory or motor impairment, or if they could not hold their head up without support. Information about the study was provided by email and informed consent was signed by the caregiver digitally (per Covid requirements). Mothers were asked to send a picture of themselves with a neutral expression before the visit. This picture was used to generate and prepare the face stimuli in advance. If a photograph was not received before the visit, a photo was taken in the lab with an iPad. Procedures were in accordance with the COVID-19 safety government regulations active at the time of data collection. The study and COVID-19 safety procedures were approved by the Department of Psychological Sciences ethics committee at Birkbeck (ref.no. 192001). Infants received a t-shirt as a thank-you.

\subsubsection{Stimulus presentation - }

Stimuli were presented on a 24 inch diagonal screen (1080p; 1920 x 1200 pixels)) and controlled by a MacBook Pro (15-inch, 2018 with a 2.6 GHz Intel Core i7 processor) using Matlab (version R2018b). Stimulus presentation was controlled with Task Engine \footnote{sites.google.com/site/taskenginedoc/} \citep{Jones2019}, Psychtoolbox 3.0.14, Gstreamer 1.14.4 for stimulus presentation, and a Lab Streaming Layer (LSL) to connect the EEG recording software to the Matlab software. A webcam (Logitech HD Pro Webcam C920) was placed on top of the screen. Open Broadcaster Software (OBS) was used to monitor the infants’ looking behaviour during the session. An iPad (7th Generation) was used for taking photographs for generating the face stimuli (see section \ref{gan}) and recording of EEG cap placement. 

The experiment consisted of a series of blocks. At the start of each block, one of the face stimuli was selected to be sampled. This stimulus was repeatedly presented for a total of 12 trials per block; we decided to present 12 trials in each block as this number of trials was 20\% higher than the typical minimum trial number (10) used in infant Nc studies to allow for data loss \citep{Munsters2019, Gui2021}. Each block started with a red spiral to attract the infants’ attention to the screen. When the infants were looking at the screen, the face stimuli were presented. This stimulus presentation was controlled with a key press by the researcher who monitored the infants’ attention via the webcam. Each trial in the block started with a fixation cross presented on a grey screen for a duration of 1000ms for the first trial, and a jittered duration between 500 and 1000ms in subsequent 11 trials. The face stimulus was presented for 500ms. Immediately after, the next trial was presented. Whenever the infants were looking away, stimulus presentation was paused and the red spiral was presented on the screen to re-attract the infants attention to the screen. A cartoon image of an object or animal was presented at the end of each block while the real-time EEG analysis and BO were performed.

\subsubsection{EEG recording and analysis of the Nc response - }

During this procedure, EEG was continuously recorded using the Neuroelectrics Enobio with an 8-channel gel-based system (NE Neuroelectrics, Barcelona, Spain). The system was connected to the recording software Neuroelectrics NIC (v2.0.11.7, Barcelona, Spain) via wifi. CMS and DLR electrodes functioned as the system's reference. Data were recorded at a sampling rate of 500Hz. Since the Nc is most prominent at frontal-central sites \citep{Courchesne1981, Gui2021}, EEG was recorded at six channels placed at locations FC1, Fz, FC2, C1, Cz, C2 in metal electrode holders in infant-sized caps (sizes K - 42cm, or KS - 46 cm). The two remaining electrodes were placed at locations P7, and Oz and functioned as our reference electrodes for re-referencing during the real-time EEG analysis. CMS and DLR electrodes were placed on the infants’ mastoid using sticktrodes.

After the presentation of each block of stimuli, real-time EEG analysis started with reading in the markers and continuous EEG from the presented block. The continuous EEG data were filtered using an FIR digital band-pass filter from 1 to 20Hz with a Hanning window (as in \cite{Webb2011}). Data were then segmented into trials based on the marker information from -100ms to 800ms after stimulus onset. Trials were baseline-corrected using the average amplitude across the -100ms to 0ms. Time series containing artefacts were identified on a channel by trial basis. Time series containing signals exceeding a threshold of -200$\mu$V or +200$\mu$V \citep{Munsters2019}, a range of 400$\mu$V, or showing a flat signal (absolute value below 0.0001$\mu$V) were marked and excluded from further analysis. Time series from all trials and channels of interest (FC1, Fz, FC2, C1, Cz, C2) were averaged together into one ERP. For the re-referencing, time series from all trials and the channels P7 and Oz were averaged together and subtracted from the ERP. 

After the pre-processing, we extracted the Nc response from the ERP. The Nc response was defined as mean amplitude calculated across the time window from 300ms to 800ms. To measure data quality, we also calculated the percentage of artefact-free trials included in the analysed ERP and the number of trials included due to lack of threshold, range, or flat signal artefacts. Both this data quality information and the ERP waveform were then displayed for visual inspection by the researchers. 

We explored different criteria on which this decision was based with different sessions (and different infants) and both automated and user-defined decisions. The pre-processed pipeline was fully automated. However, there is always the possibility of unforeseen interfering events or poor data quality that are not picked up by the automated pipeline. Interference may arise for a mother or experimenter accidentally disturbing the infant during a block, for example by talking and pointing at the screen. The signal of the EEG furthermore varies between individual infants where a certain threshold may be effective in one infant, but the thresholding may not pick up on the artefacts in another infant. To account for these possibilities, we took a manual approach and implemented a user-defined decision after each real-time EEG block which was based on the judgement of the researcher. In this researcher-based criterion, the researcher decided whether to continue, repeat or terminate the sampling. This subjective decision was based on the visual inspection of the ERP waveform, and the percentage of data from the block included in the ERP. A block was repeated when the ERP waveform was not flat or drifting and when the percentage of data included was above 20\%.

\subsection{Bayesian optimisation for sampling across the space}
\label{bayesian}
After each block, the relevant ERP metric (here, Nc mean amplitude) was passed to the Bayesian optimisation algorithm. BO is a powerful sampling algorithm that efficiently finds extrema of unknown functions, $f(x) = y$. It does so by employing a statistical model that is fitted to the sampled values and a function that guides where to sample next based on a balance between the values’ uncertainty and the previously obtained values \citep{Brochu2010}. It is particularly useful for optimising over costly functions where it is expensive to sample any given point. This is because its balance between exploration and exploitation allows the algorithm to find the function extrema in a small number of blocks. Bayesian optimisation is composed of two main parts: the surrogate model and the acquisition function. The surrogate model is a statistical model of the unknown objective function, $f(x)$. We use a Gaussian process regressor to build the statistical model based on previously sampled values, $GP(x) = p(x|y)$ \citep{Rasmussen2018}. The covariance of the predicted function is specified by a kernel, whose hyperparameters are optimised during the fitting of the model by maximizing the log-marginal-likelihood. In our system, we use the stationary Mátern kernel with a smoothness parameter $\upsilon = 2.5$ and an added white noise term. The Mátern kernel is a commonly used stationary kernel. The white noise term will estimate the global noise level of the data. This is important as the EEG signal is inherently non-deterministic and noisy. For every block of the Bayesian optimisation, the prediction of $f(x)$ and its standard deviation are passed to the acquisition function. These are used to define where to sample next while balancing both minimizing uncertainty (i.e., exploration) and maximizing the predicted function (i.e., exploitation). There are several common acquisition functions, but our system uses Expected Improvement (EI), a function that leverages information about the expected best candidate and the uncertainty of the estimations in an exploration-exploitation trade-off \citep{Kandasamy2016}. It is given by the expectation of improvement function, $I$:
\begin{equation*}
E[I(x)] = E[\max(f(x^*)-\ y,\ 0)]
\end{equation*}
As the surrogate function follows a normal distribution, we can compute EI in closed form:
\begin{equation*}
E[I(x)] = (\mu(x) - f(x^*))\ \Phi(z) + \sigma(z)\ \phi(z),\ where\ z = \frac{\mu(x) - f(x^* + \xi}{\sigma(x)}
\end{equation*}
Where $\phi$ is the standard normal density, $\Phi$ is the standard normal distribution function and $\xi$ is a hyperparameter that controls how much the acquisition function privileges exploration over exploitation. 
Because the acquisition function requires a statistical model of the target metric across the space to guide where to sample next, we need to pre-define what the first samples of the algorithm will be (i.e., burn-ins). This approach is adapted from Automatic Machine Learning research \citep{Misir2013} and it allows relevant heuristics to be added to a model prior to being optimised by the BO algorithm. The number of burn-ins and where they sample is user-defined and should be adapted depending on the paradigm being addressed. Our Bayesian optimisation sampling algorithm is run independently using python as the programming language. Before the experiment, the user needs to define the number of burn-ins, where in the configuration space they are sampled, the maximum number of iterations run by the sampling algorithm and the level of desirable exploration ($\xi$).

In the present study, our target EEG metric was the mean Nc amplitude. We optimised for its most negative value. We defined the first 4 iterations as burn-ins to Bayesian optimisation, to try to capture an initial model of the Nc amplitude’s variation across the configuration space. The initial points sampled along the interpolation between the mother’s face and stranger’s face were in order, 100\% mother’s face, 100\% stranger’s face, 1/3 mother’s face and 2/3 mother’s face. EEG experiments in neurodevelopmental research are inherently limited regarding the number of iterations it can run for one participant in each session. Because of the short attention spans of infants, we must choose our method’s parameters to minimise the number of blocks we are willing to run. Towards this end, we defined a $\xi$ value of 0.1, which benefits exploitation of identified maxima. One other reason for choosing a more exploitative $\xi$ is the configuration space being relatively small with just one dimension of variation (i.e., mother-stranger interpolation). For larger spaces, higher values of $\xi$ that benefit explorations should be considered. We defined the maximum number of iterations as 15, which we identified would be an upper bound for how many iterations we could maintain the infants’ attention (approximately 20 minutes). In the present study, the stopping criterion was always reached before the maximum number of iterations was surpassed.

\subsection{Re-iterating steps 2 and 3 until an optimum target metric has been reached}
\label{stop}
The system is run iteratively in a closed-loop. The acquisition function defines which stimulus to display next. The stimulus is presented to the participant and the target EEG metric (here the Nc amplitude as described in section \ref{bayesian}) is collected after automatic processing of the ERPs. The acquisition function will progressively choose a stimulus to sample that is predicted to be closer to the predicted maximum until a stopping criterion is met or it has run a predetermined number of iterations. The early-stopping criterion objective is to finish the program if the BO algorithm is not capturing any new information on each block. The stopping criterion should then be a relevant proxy of the uncertainty present in the statistical model. This could be the mean standard deviation after each block, or the number of consecutive times a given stimulus has been sampled. If the sampling algorithm has identified its expected maximum (i.e., the image eliciting the strongest neural signature), then it will sample this point (i.e., present the selected image) until the predetermined number of iterations are run. In theory, increasing the number of blocks, i.e., obtaining the target EEG metric multiple times for the same stimulus, could be beneficial as it would allow the algorithm to average output across blocks and obtain a more reliable value. In practice, showing the same image to the infants would induce neural habituation and decrease the strength of the neural signal for the repeated stimulus; thus, these factors need to be balanced. To balance these considerations, in the present study we used a default stopping criterion of sampling the same image three times consecutively. If an image was sampled three times, we assume that the algorithm has converged to the unknown function’s maximum.

\section{Results}
\label{results}

Good quality EEG data for analysis was collected from four infants (mean age = 6 months 4 days, range: 5 months 7 days - 7 months 22 days). One infant was excluded due to technical issues.

Here, we report the results of the real-time analysis (\ref{results1}), a demonstration that our paradigm elicits expected effects when analysed in a traditional manner (\ref{global}), and a discussion of two factors that need to be considered in the light of data collected – the balance between exploitation and exploration (\ref{exploitation}) and habituation (\ref{habituation}). 

\subsection{Results of the real-time optimisation in infants}
\label{results1}
\begin{figure}[t]
\begin{center}
\centerline{\includegraphics[width=\textwidth]{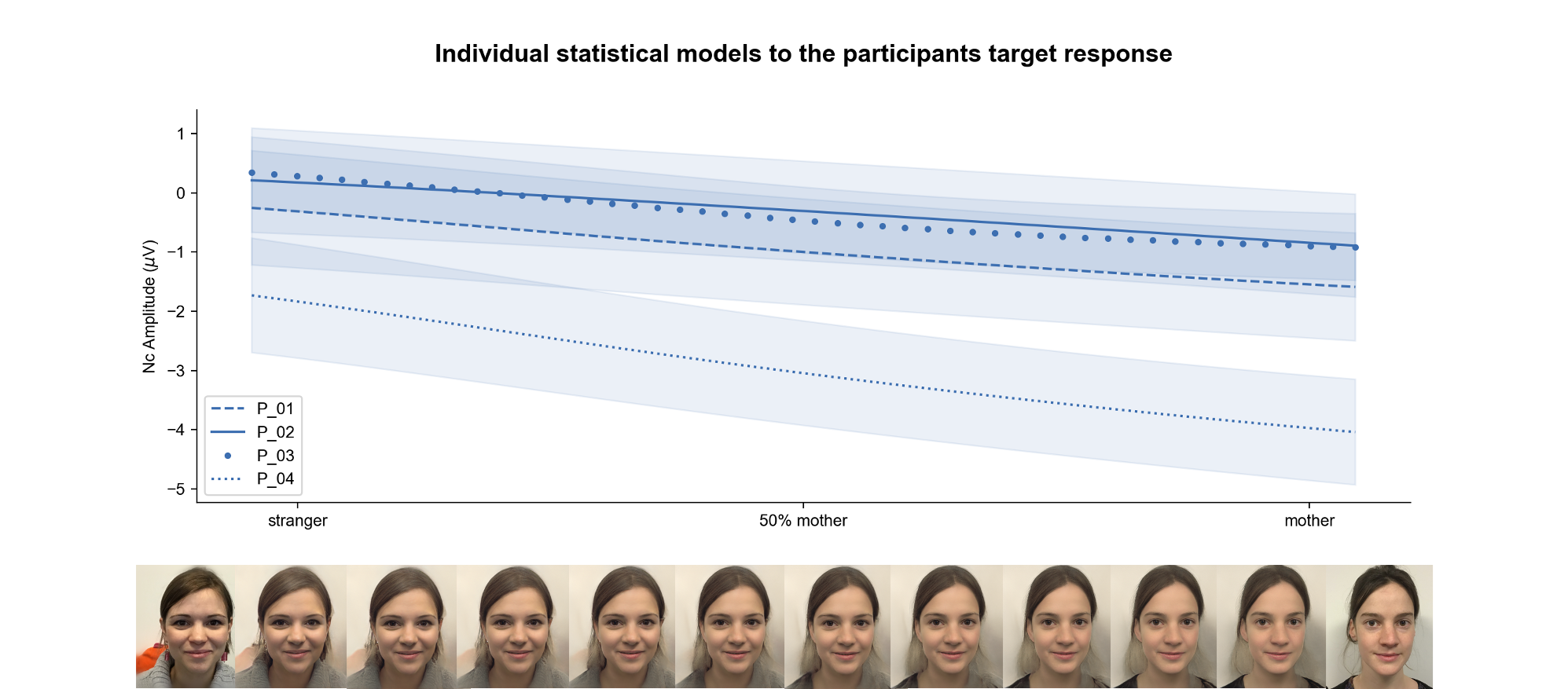}}
\caption{\textit{Individual model statistics for the four tested infants.} The BO sampling optimised the negative polarity of the Nc amplitude. The standard error for each participant’s modelled response is displayed in shaded colour around each function.} \label{individual}
\end{center}
\end{figure}

Our method allows us to build a statistical model of each participant’s response across the configuration space by fitting their sampled responses with a Gaussian Process. In the problem presented here, our configuration space was a one-dimensional interpolation between the face of the infant’s mother and a stranger. We optimised towards the largest, i.e., most negative amplitude of the Nc response (i.e., the minimum value obtained for the Nc). For all four participants the stopping criterion was met with a mean of 8.25 iterations and 1.64 of standard deviation, well below the number of possible stimuli. Furthermore, some images were sampled more than once, to allow the statistical model to assess if they indeed provided a larger Nc, as more samples of a given stimulus allow for a better estimate of the real elicited ERP.  The results are displayed in Figure \ref{individual}. As hypothesised, the predicted measures of the Nc across the stimulus continuum show a negative slope for all four participants optimised towards the most negative Nc. Thus, the statistical model predicted that for these four participants, the image of their mother would produce the most negative Nc. Because Gaussian Processes inherit the properties of normal distributions, our method returns the standard deviation of the modelled function along the configuration space. This value works as a proxy of the model uncertainty. The high values of standard deviations obtained for our modelled responses are a consequence of brain metrics not being deterministic and being highly variant even when averaging across several trials.

\subsection{Confirmation of expected mother-stranger effects using a traditional analysis }
\label{global}
\begin{figure}[h]
\begin{center}
\centerline{\includegraphics[width=\textwidth]{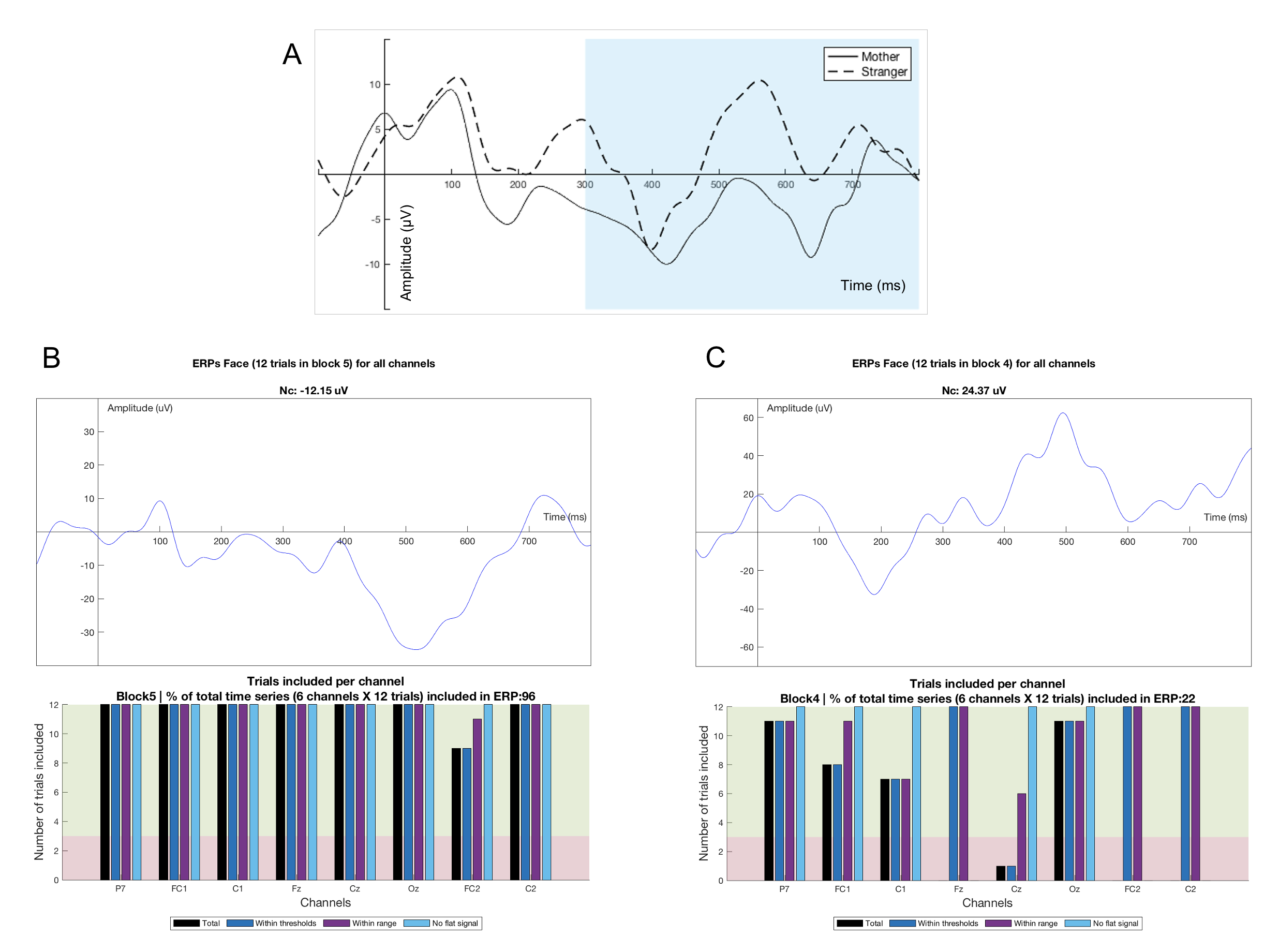}}
\caption{The grand average for mother and stranger across the four infants with the shaded area reflecting the Nc time window of interest (A). ERP and data quality report for a good ERP (B) and poor-quality ERP (C) with the ERP for the block in the top panel and the number of trials included for each channel in the bottom panel.} \label{eegresults}
\end{center}
\end{figure}

The grand average across four infants for the first two burn-in blocks (the mother and the stranger) are displayed in Figure \ref{eegresults}A. As expected, the neural response to the mother’s face showed a more negative deflection than the response to the stranger’s face, most prominent during our time window of interest (300 - 800ms). 

\textit{Illustration of Real-time metrics:}  Figure \ref{eegresults}B shows an example of the data visualisation display reviewed by the investigator after each block during data collection: an ERP with good data quality computed in real-time during data collection. In the ERP panel (top), there is a negative deflection during our time window of interest. The title of the ERP panel displays the value of the extracted ERP feature: Nc amplitude. The channel feedback panel (bottom) shows that the EEG signal, or time series, for all trials and most channels were included in the calculation of the ERP: a) the percentage of inclusion of time series for the channels of interest printed at the top is 96\%, and b) the height of bars for channel FC2 indicates that 9 time series from this channel were without any artefacts (black bars), 9 were within the thresholds (blue bars), 11 within the range (purple bars), and 12 did not display a flat signal (light blue bars). The bars for the other channels indicate that all 12 time series for the other channels were without any artefacts (thus, within the thresholds, within the range, and without a flat signal). Due to the high amount of clean time series, and a clear ERP waveform, the researchers decided to continue sampling after the data collection of this block. 

Figure \ref{eegresults}C displays the real-time EEG feedback for a low-quality ERP for one of the infants that was excluded due to flat channels and excessive movement. In this low-quality ERP, the waveform does not show the typical shape and shows a positive deflection rather than a negative deflection during our time window of interest (ERP panel, top). More importantly, the channel feedback panel (bottom) shows that 22\% of the time series from the channels of interest were excluded in the calculation of the ERP. For both channels P7 and Oz, 11 out of the 12 time series are included in the calculation of the ERP. This suggests that the EEG data from the channels used as reference are relatively good quality. For channels Fz, FC2, and C2, there were no artefact-free trials (absence of black bars) due to flat signals (absence of light blue bars, whereas the height for the ‘within threshold/range’ bars are 12). For channels FC1, C1 and Cz, not all 12 presented trials were included either. This was mainly caused by the time series exceeding the thresholds and/or the range. In instances like these, the researcher would decide to repeat the block due to the low percentage of included time series and the low-quality ERP waveform. The researchers would first attempt to improve the EEG signal for the channels with low data quality (here, FC1, C1, Fz, Cz, FC2, and C2) by adjusting the cap and/or regelling the EEG electrodes before repeating the block. 

\subsection{The exploration vs exploitation parameter}
\label{exploitation}
As mentioned in section \ref{bayesian}, the acquisition function, that guides the sampling algorithm, contains a hyperparameter $\xi$ that controls the level of exploration of the space. In our proof-of-concept we chose a conservative value of 0.1, benefiting exploitation, as the number of blocks we can run with infant participants is severely limited. Figure \ref{exploitationresults}A shows the surrogate model and the acquisition function for one of the participants after 4 burn-ins. The surrogate model’s mean is a proxy for how the model interprets the Nc amplitude to change across the space and is displayed with a dashed line. The standard deviation of the surrogate model is interpreted as the uncertainty value across the fitted space. The uncertainty is minimal for the points already sampled and it increases the further it is from these points. This is due to the prior assumption that stimuli closer to each other in the configuration space elicit a similar response of the Nc ERP. The different acquisition function plots display the effect of different values of $\xi$ for the next point to sample for this given participant – represented with a star. The higher the $\xi$ parameter, the more exploratory the sampling behaviour and the more it will investigate sampling the regions where uncertainty is highest. At the extreme, the acquisition function will display active learning properties, sampling only to minimise the standard deviation of the space. The lower the $\xi$ value, the more it will look into resampling what it found the highest value to be, disregarding uncertainty when $\xi$ is 0. The small number of iterations required to achieve the stopping criterion and the fact that the model captured the mother-stranger variation in Nc amplitude are good indications that a conservative value was the right choice for this specific paradigm.

\subsection{The problem of habituation}
\label{habituation}
\begin{figure}[h]
\begin{center}
\centerline{\includegraphics[width=\textwidth]{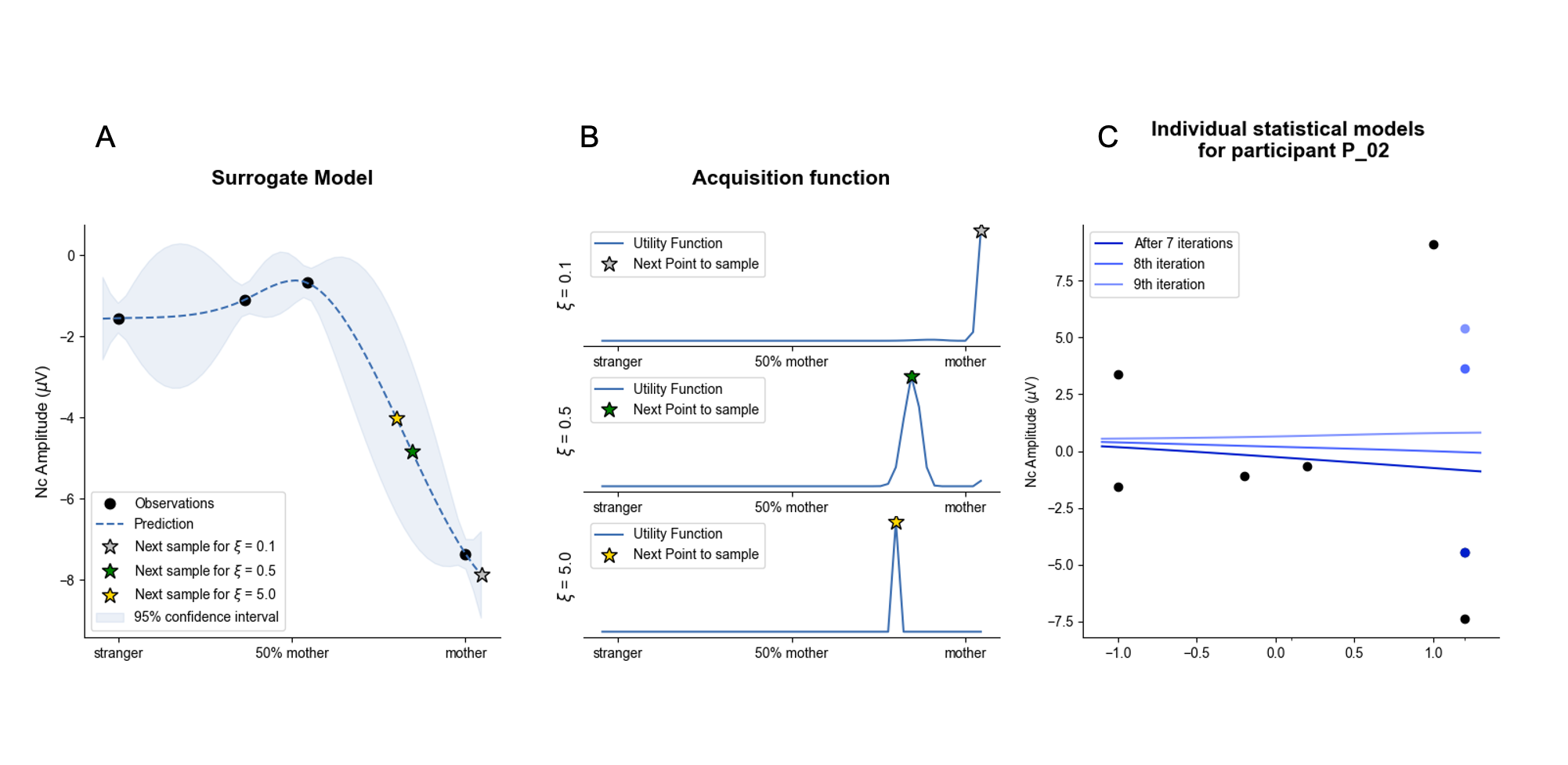}}
\caption{\textit{Case-study of exploration, exploitation and habituation on participant P\_02.}
Model statistics for participant P\_02 after four iterations with the next points to sample from three different utility functions marked as stars (a). Acquisition function for the model statistics presented in (A) for three different values of $\xi$ (B). The larger the value the more exploratory is the utility function and more will it privilege uncertainty over sampled maxima. The point to sample next is marked with a star. Example of habituation on participant P\_02, where positive amplitudes of the Nc result in a change in the model statistics (C).
} \label{exploitationresults}
\end{center}
\end{figure}

Efficient sampling in EEG research and, more specifically in neurodevelopmental research, is fundamental to minimise problems of habituation or of short attention spans that plague the field \citep{Snyder2008}. This makes the case for algorithms that can predict individual response to a collection of stimuli, while only sampling a subset of them. We show in Figure \ref{exploitationresults}C that, even while minimizing the number of iterations, there are signs of habituation to the stimuli on P\_02 that will shift how the model statistics of the space are predicted. When trying to predict an individual response to a set of stimuli, it is imperative to try to minimise signal variations that aren’t directly related to the stimulus being displayed. To avoid the effect of the habituation to the stimulus on the model statistics for a given participant, we considered, for the proof-of-concept, the predicted model statistics before any habituation to the repeated stimulus had occurred for participant P\_02 in our post hoc analysis.

\section{Discussion}
In this work we presented neuroadaptive EEG, a method that uses real-time data processing and machine learning algorithms to invert common research approaches by searching a large stimulus space to find the stimulus that maximally evokes a given neural response. In a proof-of-principle study, we show that neuroadaptive measurement of infant brain activity can locate a picture of the infant’s mother from a one-dimensional face space. Here, we discuss the advantages and limitations of this method; technical considerations for its use; and the potential of its application to a broad range of research questions across diverse fields. 

\underline{\textit{Scientific robustness}}

Neuroadaptive EEG involves a fully closed-loop design. Thus, the neural feature targeted in the study and the stimulus search space must be predefined and hardcoded into the experiment itself. This completely removes analytic flexibility from the investigator since data is analysed during the experiment itself, solving one of the major challenges of current neuroimaging research \citep{Nosek2015}. Further, our approach requires brain signals to be reliable on an individual level within the experiment itself (if not, the search will not converge). This approach can be combined with other advances in robustness, such as external pre-registration of the selected EEG features and selected stimulus space to avoid the ‘file drawer’ problem. The approach is thus most suitable when the investigator has a known brain metric they are interested in investigating. However, the range of potential metrics to which the method can be applied is very broad; this could include connectivity between particular regions or at particular frequencies; activation in particular brain locations; or the speed or amplitude of well characterised event-related responses. In this way, the experimenter can build on the long history of stimulus-driven investigation of particular brain metrics whilst significantly extending our understanding by mapping new stimulus landscapes. Of note, this differentiates our approach from more traditional brain-computer interface approaches (BCI) where the target metric is data-derived for each individual, making it hard to use as a tool for cumulative discovery. Finally, neuroadaptive methods support effective generalisation by computing the modulation of a metric across a large stimulus space. The importance of considering the limitations of inference to the specific stimuli selected in any given paradigm has recently been elegantly outlined \citep{Yarkoni2017a}. With neuroadaptive optimisation, the boundaries of generalisation can be objectively and efficiently probed. This approach is not only applicable to neuroscientific research as it can be used more broadly to multiverse analysis in general \citep{Dafflon2020}.

\underline{\textit{Efficiency of estimation across a space}}

Our proof-of-principle case was the use of these methods in infants. Infants are a particularly challenging population because of their limited attention spans and inability to respond to verbal instruction, and this makes it fundamental to be able to capture dependencies between stimuli using as few iterations as possible. In this experiment, we use a one-dimensional configuration space that interpolates between the stimulus of the participant’s mother face and a stranger’s face. We measure the mean amplitude of the Nc ERP as the target EEG metric. Contrary to classic research paradigms, our method does not restrict us to just two stimuli (i.e., mother and stranger), but we are able to capture the continuous variation between these two images. This opens the possibility to explore not only if there are differences in the ERP amplitude, but also whether the signal variation is gradual or if it is abrupt, such that only the last image of the mother elicits a stronger Nc (as may be the case if face perception is categorical \citep{Leopold2010, Moulson2011}. Here we present the individual model prediction across the configuration space to show individual variation, but our method allows for the use of other data to account for intra-group variability; for example, the measure of uncertainty across the space can be useful to capture differences in signal reliability between participants. The path of exploration to exploitation used by the algorithm can also be relevant to discriminate between participants. Our method showed this capability in our proof-of-concept by predicting the variation of the signal across the stimuli while only sparsely sampling them. This method can be used in future neurodevelopmental research to test the reliability of individual ERPs as correlates of cognitive functions and to understand the commonality and individual differences in neural responses to visual stimuli.

\underline{\textit{Diverse populations}}

Neuroscience has suffered from a long history of collecting data in primarily White, neurotypical, English-speaking young adults, often from higher-education institutions. The need to expand to broader and more diverse populations is well-rehearsed but comes with significant challenges. Beyond the practicalities, we must recognise that selection of stimuli based on theories derived from our existing narrow samples may not be the most appropriate or efficient way to learn about brain function in more diverse cultures. Such an approach is highly prone to cultural bias (e.g., selection of White faces for experiments with diverse ethnicities) and a deficit-based perspective (identifying differences in how neurodivergent adults respond to stimuli selected by normative experimenters – such as examining diminished responses to faces in autism).  Our approach allows the investigator to move away from studying brain responses to the same stimulus in different groups, and towards studying how a comparable brain metric is modulated across a much broader stimulus space. Rather than asking why a young child with autism doesn’t attend to faces, we can search for the type of stimulus that the child does find interesting. Of course, these approaches are complementary rather than exclusive, but together they may provide us with a richer suite of tools to study brain function across diverse populations and to devise individualised interventions that build on strengths.

These considerations are particularly important in studying the infant brain, where one of our core interests is the way in which mature cognitive topography emerges in development. For decades, researchers have debated whether the mapping between brain system and cognitive function is present at birth (nativist or modular accounts \citep{Markram2006} or whether brain regions have functions that change over developmental time \citep{Johnson2011}. One leading account proposes that brain regions progressively specialise through a process of interaction and competition \citep{Johnson2011}. Testing such accounts is slow and challenging with traditional experimental paradigms, because stimulus selection is typically informed by adult cognitive models (e.g. since the fusiform face area or N170 component are face-selective in adults, we examine their response to faces vs objects in infants \citep{Deen2017}. With a neuroadaptive approach, in principle investigators can move to map the modulation of the fusiform face area or N170 across a much broader stimulus space. Indeed, GANs can be used to create a large library of both faces and objects that are artificially manipulatable along a range of dimensions. In this way, we may generate new knowledge about the cognitive topography of the infant brain - seeing the world through their eyes. 

\underline{\textit{Methodological considerations}}

The approach includes several parameters that can be adjusted depending on the research question that is being addressed. Our method allows the researcher to navigate the exploration-exploitation boundaries of the configuration space by defining the $\xi$ value, which allows the algorithm to balance between a more exploitative search of the maximum target metric. Experimental designs that are limited to a small number of iterations (because of the population studied), as was the case of our proof-of-concept with infants, should use a low value of $\xi$ and benefit exploitation of the stimuli. Designs that explore larger configuration spaces, with many dimensions and stimuli, could utilise a larger value of $\xi$ and exploration to better capture the modelled statistics of the space. As an example, if instead of just varying along one dimension (e.g., mother-stranger interpolation), the configuration space contained several semantic variations (e.g., a 4-dimensional space with faces varying across age, emotion, gender and eye-to-eye distance), a more explorative behaviour would be beneficial.

Selecting a robust target metric is important. The algorithm searches the space to identify the stimulus that elicits the maximum target metric we are measuring. For the four participants for whom the target EEG metric for the Bayesian optimisation was the most negative amplitude of the ERP, we were able to model a variation of the signal that maximised its negative amplitude for images of the mother’s face. There are several possible explanations for this; one may be that categorical face perception means that the infant is sensitive to varying degrees of proximity to the mother, but not varying degrees of ‘strangerness’ \citep{DeHaan1999}. The success of BO-based approaches are dependent on reliable and meaningful neural signatures to be used as target metrics. In the current proof-of-concept study, we chose our parameters and measures for the ERP features based on the previous literature and offline analysis of pilot and existing datasets \citep{Gui2021}. The preprocessing methods and calculation of ERP features however vary between studies. It is possible to further examine how the process of optimisation would vary when using different EEG features or metrics. For example, in line with most previous work \citep{Webb2011, Luyster2014} we used the mean amplitude from 300 to 800ms as an EEG metric. Selecting appropriate data processing parameters is also important. Researchers willing to use this method should conduct some preliminary offline analyses of individual-level ERPs to examine how varying recording and preprocessing methods affect the quality of the ERP waveform. Different thresholds or a larger number of trials within a block may improve the quality of the ERP waveform and decrease the signal to noise ratio. In our experience, with infant data increasing the duration of the experiment and exposure to the same image within a block by including more trials does not improve the quality of the ERPs. Inspections of the individual infant ERPs obtained with a varying number of trials using existing and pilot datasets confirmed that the choice of 12 trials was sufficient for our experiment. Another avenue is to develop an EEG metric that reflects the quality of the ERP waveform, such as standard deviation or area under the curve during the baseline, or a quantification of the shape of the ERP waveform and curve. One could implement this information into the BO algorithm by adding varying weights of the samples. Given that the present study aimed to test that BO could appropriately map neural signatures of attention engagement, we reasoned it was essential to define a threshold for inclusion of good quality trials per block and channel. In case of poor EEG quality data we preferred to exclude the block completely and repeat the data collection for the same image. Here, the decision on how to proceed with the iterative optimisation process was dependent on the subjective inspection of the researchers. This always raises the risk of potential bias, particularly given the ERP and key metrics were displayed- although this was done to allow inspection of data quality, it raises the possibility that researchers could be biased towards rejection or selection based on the nature of the Nc response.  Implementing and quantifying data quality measures would make this process more objective and therefore less potentially biased; however, in our pilot work automated quality measures weren’t considered as effective at detecting poor data quality as a human researcher. Further work in building stronger automated quality measures for the signal are needed. 

\underline{\textit{Limitations}}

The sampling algorithm and the real-time EEG can work with any configuration space setting, but our configuration space builder is limited to 16 pre-defined meaningful variations of faces as described in section \ref{gan} (e.g., pitch change, age). Other meaningful variations of faces can be learned by following the approach described in section \ref{gan}. Non-facial stimuli are currently not supported by the configuration builder but could be integrated in future work. The sampling algorithm is further limited by the quality of the target EEG metric. Despite the algorithm being flexible to receive noisy input, if the averaging of the EEG signal per block fails to capture the dynamic of the ERP, the statistical model will not be able to capture the ERP dependency across the space. It is important to note that we are not proposing that the neuroadaptive method should replace traditional stimulus-driven approaches. Traditional methods will be important to further the discovery of new neuroimaging metrics of interest. However, once such metrics have been identified and sufficiently well parameterised the neuroadaptive method allows the investigator to map the modulation of these metrics across a larger stimulus space, providing a complementary tool to further our understanding of the cognitive topography of the brain.

\section{Conclusion}
A core goal of functional neuroimaging is to study how the environment is processed and represented in the brain. The mainstream paradigm involves concurrently measuring a broad spectrum of brain responses to a small, preselected set of environmental features selected with reference to previous studies or a theoretical framework. As a complement, we invert this approach by allowing the investigator to record the modulation of a preselected brain response by a broad spectrum of environmental features. We show that online recording of the Nc infant brain engagement response can automatically identify the position of an individually salient face (the infant’s mother) in a one-dimensional face space. We demonstrate the promise of this approach for studying the developing brain, where our theories based on adult brain function may fundamentally misrepresent the topography of infant cognition and where there are substantial practical challenges to data acquisition. Our approach may also have significant potential in areas where theoretical frameworks or previous empirical data are impoverished or misleading, allowing us to tackle new questions and move beyond heteronormative undergraduate student populations. Furthermore, by using a prespecified closed-loop design the approach tackles fundamental challenges of reproducibility and generalisability in brain research. We believe this approach has substantial potential in infancy research and beyond for accelerating our understanding of the cognitive topography of the brain.

%
%
%

\ACKNOWLEDGMENT{We acknowledge Laura Carnevali, Teresa Del Bianco and Jannath Begum Ali for help with the study set-up. This work was supported by the European Union’s HORIZON 2020 Research and Innovation Programme under the Marie Sklodowska-Curie Grant Agreement No 814302. This study received funding from the Economic and Social Research Council grant n. ES/R009368/1. This work was supported by EU-AIMS (European Autism Interventions), which received support from the Innovative Medicines Initiative Joint Undertaking under grant agreement no. 115300, the resources of which are composed of financial contributions from the European Union’s Seventh Framework Programme (grant FP7/2007-2013), from the European Federation of Pharmaceutical Industries and Associations companies’ in-kind contributions, and from Autism Speaks as well as AIMS-2-TRIALS which received support from the Innovative Medicines Initiative 2 Joint Undertaking under grant agreement No 777394. This joint undertaking receives support from the European Union’s Horizon 2020 research and innovation programme and EFPIA and AUTISM SPEAKS, Autistica, SFARI (RH, EJ). Disclaimer: The views expressed are those of the authors and not necessarily of the IMI 2JU. }





\bibliographystyle{informs2014} 
\bibliography{neuroadaptive_eeg.bib}

\end{document}